# Generalized Resemblance Theory of Evidence: a Proposal for Precision/Personalized Evidence-Based Medicine

Maani Beigy [1,2,3] *

[1] Department of Epidemiology and Biostatistics, School of Public Health, Tehran University of Medical Sciences, Tehran, Iran

[2] Students' Scientific Research Center (SSRC), Tehran University of Medical Sciences, Tehran, Iran

[3] Rheumatology Research Center, Tehran University of Medical Sciences, Tehran, Iran

November 2015

**Abstract**

Precision medicine emerges as the most important contemporary paradigm shift of medical practice but has several challenges in evidence formation and implementation for clinical practice. Precision/Personalized evidence-based medicine (pEBM) requires theoretical support for decision making and information management. This study aims to provide the required methodological framework. Generalized Resemblance Theory of Evidence mainly rests upon Generalized Theory of Uncertainty which manages information as generalized constraints rather than limited statistical data, and also Prototype Resemblance Theory of Disease which defines diseases/conditions when there is a similarity relationship with prototypes (best examples of the disease). The proposed theory explains that precisely-personalized structure of evidence is formed as a generalized constraint on particular research questions, where the constraining relation deals with averaged effect sizes of studies and its comparison to null hypothesis; which might be of either probabilistic or possibilistic nature. Similarity measures were employed to deal with comparisons of high-dimensional characteristics. Real examples of a meta-analysis and its clinical application are provided. This is one of the first attempts for introducing a framework in medicine, which provides optimal balance between generalizability of formed evidence and homogeneity of studied populations.

**Abbreviations**: EBM, evidence-based medicine; pEBM, personalized/precision evidence-based medicine; IPD, Individual Participant Data; PrM, precision medicine; PM, personalized medicine

* Email: m-beigy@student.tums.ac.ir; manibeygi@gmail.com

# 1. Introduction

Over the past 25 years, evidence-based medicine (EBM) has been developed as a modern medical paradigm trying to integrate the *individual medical expertise* and *external scientific evidence* to provide the best available evidence. Despite its revolutionary effects, this paradigm of modern medicine has major imperfections so that EBM is still a disease-oriented paradigm rather than a patient-oriented (i.e. personalized) manner (Bensing 2000). The suitable information is extracted from randomized clinical trials (RCTs) (or meta-analysis and systematic reviews) and stems from studies with tightly formulated inclusion/exclusion criteria which are used for studying eligible patients in RCTs. These tight criteria eliminate common comorbidities and other characteristics (covariates) of participants to increase "statistical power" of detecting the *causal* association (Bensing 2000). Unfortunately, only 5 to 10% of all eligible adults are engaged in clinical trials (Gelijns and Gabriel 2012). The question now arises why we have to rely on the outcomes extracted from *disease-oriented evidence (DOE)* while our patients are not highly *similar* to the studied populations of DOE. Also, some may inquire into why *POEM (Patient-Oriented Evidence that Matters)* (Smith 2002) studies undergone less attention, despite their well-known necessities in medicine. Due to these imperfections, the process of generating medical evidence is slow and inefficient (Ma et al. 2015). Based on one analysis on the results of Cochrane Systematic reviews (Dib et al. 2007), only about 42% of interventions were likely to be beneficial; about 7% of interventions were likely to be harmful, and in 49% of reviews insufficient evidence was reported. Systematic reviews and meta-analyses are designed to reduce uncertainty in health-care decision making, but we encounter bitter facts about them. For example, in the chapter of breast cancer in *Harrison's principles of internal medicine* we can see this confession: "meta-analyses have helped to define broad limits for therapy but do not help in choosing *optimal* regimens or in choosing a regimen for *certain subgroups* of patients" (**Longo** et al. 2012).

To date, the proposed solutions have not yet overcome those challenges of non-precise/non-personalized medical practice. Some may say these efforts are at the beginning of their way, but I believe ill-defined paradigms with vague objectives and methods are not fruitful for these purposes. Three terms that are often used interchangeably have been employed to explain our ideal practice, which are somewhat similar but with many misunderstandings. The first one is *patient-centered medicine* (PCM) which has basically a humanistic, biopsychosocial perspective, combining ethical values on "the ideal physician", with psychotherapeutic theories on facilitating patients' disclosure of real worries, and negotiation theories on decision making (Bensing 2000). It majorly focuses on the participation of patients in clinical decision making by considering the patients' perspective and adjusting medical care to the patients' needs and preferences (Bensing 2000). The second paradigm for modern practice is *personalized medicine (PM)* in which diagnostic tests, mainly genetic information, are used by physicians "to determine which medical treatments will work best for each patient" (2014). Thus, by employing the data from those genetic tests and *individuals' medical history*, and considering patients' circumstances and values, *individually-targeted* options for treatment and prevention might be achievable (Hamburg and Collins 2010; PMC 2014). The third term, which is recently developed, is *precision medicine (PrM)* in which treatments are targeted to the needs of individuals with respect to individualized information of genetic, biomarker, phenotypic, or psychosocial characteristics; in order to empower physicians to distinguish a given patient from others (Jameson and Longo 2015). From the philosophic standpoint, these efforts have not yet made the paradigm shift we need. It seems there is a long way to the goal of having precisely-personalized, patient-centered *AND* evidence-based medical practice; which may mainly be attributed to our inefficient models of *evidence formation* and the failure to recognize medical research perceptions e.g. causality, confounding, interaction, etc. What we need is a *clinician-friendly* framework which facilitates the process of personalized evidence formation, but not the *models* with high computational complexities and limited to analysis of *collective (aggregate)* data. It is clear that further than traditional randomized clinical trials and meta-analyzes, studying the individual participant data (IPD) (Tierney et al. 2015) and also the results of daily practice is necessary. In the other words by using radically different inferential methods, those characteristics that are being presently considered disturbing and distasteful in the reasoning process are desired to be recognized and properly treated in personalized studies.

Uncertainty does have many aspects in medicine, which needs to be addressed for the evidence theory of personalized/precision EBM (pEBM). Sources of information uncertainty (Coppi et al. 2006) in medicine might be summarized as:

1) Uncertainty attributed to the link between the observed data (of samples) and the *universe* of possible data (populations). It mainly addresses how much the recruited individuals are representative of the reference population.

2) Imprecision in measuring the empirical phenomena.

3) The vagueness connected with the use of natural language (NL) in describing the real (i.e. clinical) world.

4) Neglecting the theoretical assumptions (e.g. normality, adequacy of sample size) and specific observational instances of values (e.g. missing data, extreme values).

5) Imprecision in the *granularity* of the NL terms and concepts (Zadeh 2006).

6) Heterogeneity of recruited individuals (e.g. variance, dissimilarities) and subgroups of diseases, mainly due to the high-dimensional characteristics (genetic, biomarker, phenotypic, or psychosocial) (Ma et al. 2015; Wang et al. 2007).

7) Generalizability of evidence for particular individuals in clinical settings.

As a primordial principle, our medical practice has to be *precisely-personalized* in terms not only "to determine which medical treatments will work best for each patient" but also "to precisely determine the degrees of possibility that a given treatment causes specific effects for a given patient with determined genetic, biomarker, phenotypic, and psychosocial characteristics".

To begin with methods section, I represent seven fundamentals of Generalized Resemblance Theory of Evidence and also the employed theories and concepts. What follows in results is a précis of my proposal, which is elucidated in discussion.

## 2. Methods

### 2.1. Seven Fundamentals of Generalized Resemblance Theory of Evidence

*Fundamentum 1.* **Medical theories *change* as science *develops*, experiments might be *confirmed*, but medical evidence is *formed***

To reconstruct the backbone of the evidence theory of pEBM, I have to rephrase and clarify the previously defined philosophical standpoints toward scientific development of modern medicine. I believe, just the same as Kazem Sadegh-Zadeh elucidated (Sadegh-Zadeh 2001), science development, especially in medicine, has a deductive pathway in which the progress is obtained when the theoretical meanings, perceptions, concepts, and informatics paradigm of scientific methodology is developed. It means despite the common sense, the progress of medical knowledge has not been achieved cumulatively, though its progresses has been occurred after initiatives of developed or changed *major* theories (fig 1).

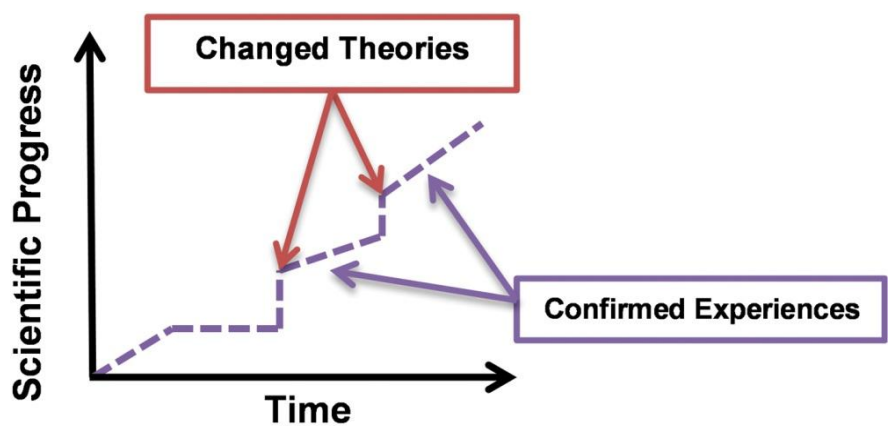

Fig 1 A tentative illustration of medical scientific progress with respect to changed major theories and confirmed experiments. It elucidates that both confirmed experiments and changed theories are crucial for progress in medical knowledge; however, the impact of changed meanings, perceptions, concepts, and informatics paradigm is paramount.

There are lots of examples indicating the initiation of a paradigm shift before formation of problem-oriented evidence; one of the recent examples might be our view to cellular differentiation in regenerative medicine research in which we had previously believed that it is impossible to reproduce undifferentiated pluripotent stem cells from mature differentiated cells (Okita et al. 2007; Yu et al. 2007). After the revolution in our standpoint toward stem cells and our view to cell-tissue development; the currently developing body of evidence regarding regenerative medicine has become achievable, so that novel therapeutic approaches are going to be introduced in medical literature. Altogether we can reasonably summarize the process of scientific development of medicine as a form of discontinuous evolution in which the preceding theory of $T_i$ is replaced by theory $T_j$ (Sadegh-Zadeh 2001). It is particular to note this statement does not indicate $T_i$ is necessarily false and/or $T_j$ is demonstrated to be true, but because of developing and changing those previous meanings, perceptions, concepts, and informatics paradigm; the revolutionary theory $T_j$ is employed to construct the required problem-oriented evidence (Sadegh-Zadeh 2001).

The other cornerstone of scientific development of modern medicine is the concept of *confirmed experiments*. It is clear that medical knowledge mostly belongs to experimental science, so the majority of medical knowledge is obtained by replication studies. It is particular to note that medical experiments (i.e. medical studies in the realm of EBM) are not exactly-repeatable because of variable nature and time varying uncertainties of biologic systems. However, each study culminates in some information which is a part of a greater concept we call evidence. Hence, this information have to, and are potentially able to, be confirmed by other replication studies. Here, replication does not mean the experiments have to be conducted with maximal similarity to the initial studies in terms of methodology and design, but replication studies do share certain well-defined objectives.

Clinical guidelines and other practical information that we, clinicians, use as evidence to do our daily practice is the fruit of a longtime process that I call here *evidence formation*. Therefore, evidence formation indicates more than conclusions of medical experiments; more exactly it tries to model the process of gathering *precise* practical sentences which clinicians use; for instance: "insulin is a good first-line therapeutic choice for all patients with type 2 diabetes mellitus, especially for patients presenting with A1C >10%, fasting blood glucose >250 mg/dl, random glucose consistently >300 mg/dl, or ketonuria (McCulloch et al. 2013)". I want to clarify that evidence formation is conducted through several confirmatory studies/experiments, and is terminated when sufficient information has been obtained to enable us to write a precise sentence in terms of clinical usage.

*Fundamentum 2.* **The Logic of Medicine is multi-valued (non-classical), because its fundamental concepts (e.g. health and disease, etiology, etc.) are Meinongian objects**.

It is previously demonstrated that health, illness, disease, etiology, bias, confounding, interaction and other fundamental concepts of medicine are non-Aristotelian attributes because they violate the principles of excluded middle and non-contradiction (Sadegh-zadeh 2000); thus these concepts are considered both *impossible* and *incomplete* objects based on Meinong's theory of objects (Reicher ; Sadegh-zadeh 2000), we have to treat them properly (fig 2). It is clear that logical system of medicine is not consistent with Classical Logic mainly because of the non-homogeneities and Meinongian nature of medicine (Sadegh-zadeh 2000); therefore, multi-valued logics such as Fuzzy Logic or precise probabilistic inferential systems (e.g. Dempster-Shafer three-valued probabilistic reasoning) seem to better handle uncertainties of medical evidence formation. However, it should be differentiated from routine clinical reasoning which is somewhat more abstract than the evidence formation process, so that the computational complexities of the aforementioned logics are not easily handled in daily activities of clinicians, thus we might reasonably consider the process of clinical practice to be managed in a deontic logic environment (Sadegh-zadeh 2000). I believe my proposal for theory of evidence might be easily translated for a deontic environment because of its dependency to natural language analysis.

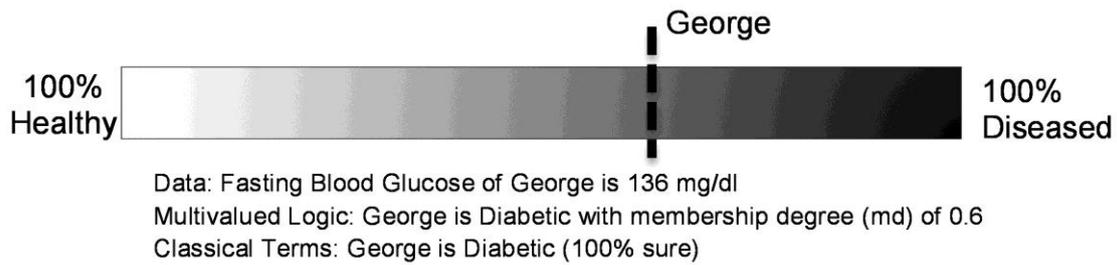

Fig 2 The representation of Fuzziness of the concepts of health and disease.

*Fundamentum 3*. **Evidence is formed mainly by the information obtained from four cardinal sources of heterogeneity in populations (doomed, resistant, at-risk, or protected).**

One of the other important concepts of medicine which, unfortunately, is still ill-defined is heterogeneity. To date, variance (in statistical standpoint) and dissimilarity have been considered as the sources of information heterogeneity of medical evidence so that in meta-analyses the *effect sizes* (of given outcomes e.g. getting diabetes) of different studies are usually weighted by *inverse variance* (along with a correction constant for heterogeneity in random-effect model) to be combined as *averaged effect size* (Higgins and Thompson 2002). Nevertheless, our primary purpose to measure heterogeneity is to decide whether these studies are homogeneous enough (in terms of characteristics of study populations e.g. clinical manifestations, genetic background, inclusion/exclusion criteria of studies, conducted methods etc.) to be combined for evidence formation. The latter source, which is even more important, is taken into account by, solely, qualitative checklists (Manchikanti et al. 2008). In epidemiology, four cardinal destinies occur (Massad et al. 2003). 1) Doomed people: they would develop a given disease, independently of having exposure to certain risk-factors/interventions or having the suspected cause. 2) Resistant people: they never develop the disease, regardless of being exposed or not. 3) Protected individuals develop disease when they are not exposed thus exposure is a protection factor. 4) At-risk individuals develop the disease only if exposed to the suspected cause. The effect sizes of studies (e.g. odds ratio, relative risks, number need to treat, etc.) are obtained by calculations on probabilities/possibilities of membership to each of these subgroups. Since evidence is mainly formed by these effect sizes, it is necessary to determine heterogeneity (variance, similarity, etc.) of people pertaining to these subgroups (fig 3).

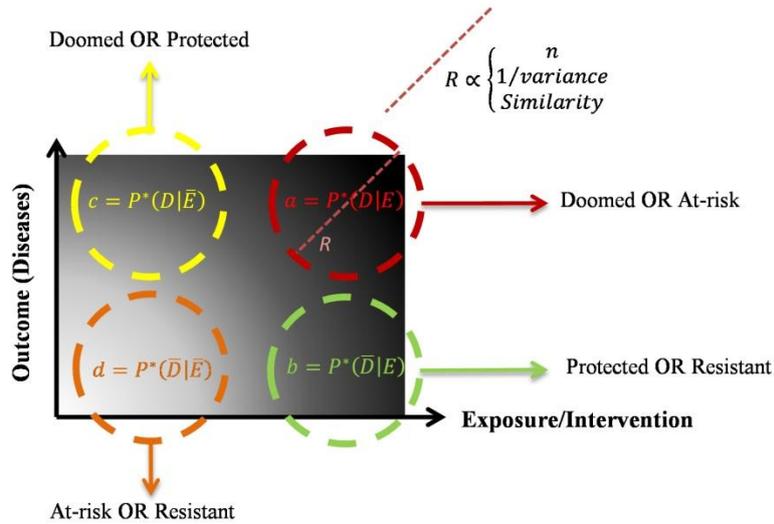

Fig 3 Four cardinal sources of heterogeneity in populations (doomed, resistant, at-risk, or protected) with respect to exposure of individuals to interventions or risk-factors are shown. $P^*(D|E)$ stands for the conditional probability or conditional possibility (membership degree) of developing a given disease ($D$) if exposed to a certain risk-factor or intervention ($E$). These individuals might be *at-risk* or *doomed*. The probability/possibility of being in the second group, is obtained by $P^*(\bar{D}|E)$ conditional probability/possibility of not developing the given disease ($\bar{D}$) if exposed to the suspected risk-factor or intervention ($E$), which might be *protected* or *resistant*. The individuals of the third group might be *doomed* or *protected*, in which $P^*(D|\bar{E})$ is the conditional probability/possibility of getting the disease D if not exposed to the suspected cause ($\bar{E}$). Finally, the conditional probability/possibility of being *at-risk* or *resistant* is obtained by $P^*(\bar{D}|\bar{E})$.

*Fundamentum 4.* **Evidence formation starts from a singularity then propagates a core of data i.e. the initial information about a question**

When there is a problem-oriented (i.e. clinical) research question, and consequently a hypothesis, but there is no study addressing that hypothesis, there is maximal vagueness and uncertainty which might be called *evidence singularity* (fig 4). We overcome this vagueness by conducting an initiative study which culminates in some data. The conclusions driven from this preliminary data are then propagated by further studies. For instance, our question is "do pre-diabetic individuals convert to diabetes without intervention?"; the preliminary data might show that "more than 70% of pre-diabetic individuals develop type 2 diabetes mellitus during 5 years". This *fundamentum* indicates evidence formation starts with a precise question, and requires an initial core of information to achieve precise practical sentences of clinical usage and productive evidence formation.

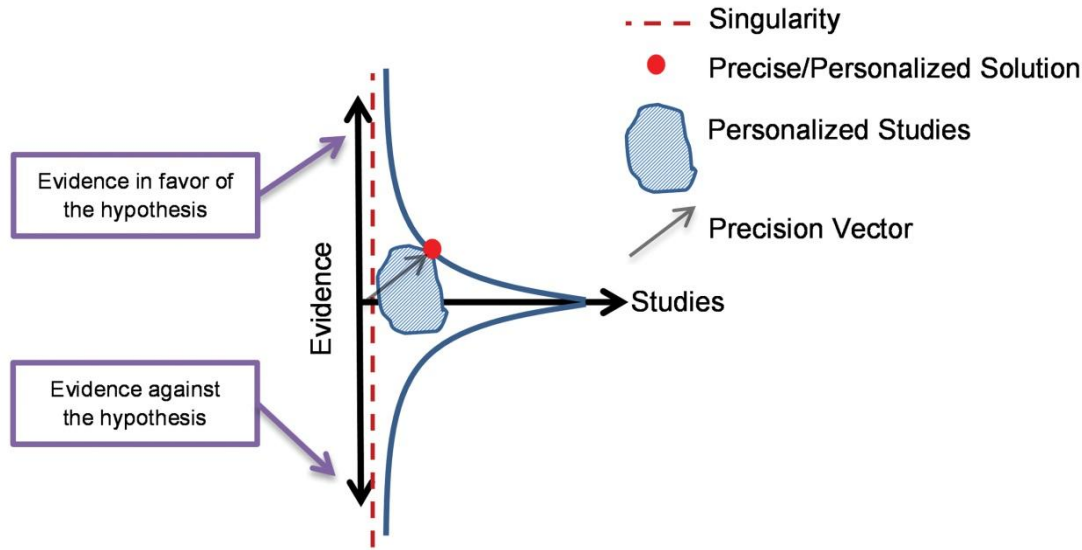

Fig 4 A tentative illustration of *evidence formation* with respect to the initial question/hypothesis and the studies addressing that issue. Here, when there is a practical question but no studies about that, we have *evidence singularity* which is vanished when some studies provide the personalized/precision information. The personalized solution for a particular patient is obtained by meta-analyzing those studies, and is determined when the characteristics of a vector (named *precision* vector) are addressed.

*Fundamentum* **5. Evidence formation culminates in sentence(s) of precisiated natural-language (PNL).**

As mentioned in *fundamentum* 1, the final conclusions of evidence formation are translated into sentences of NL, which is based on linguistic variables/functions of medical terminology. First, it should be noted that based on *fundamentum 2* all medical conditions are, or are allowed to be, *graduated*, which means, "be a matter of degree or, more or less equivalently, fuzzy (Zadeh 2006)". Moreover, all variables of medical research are granulated, with a granule defined "as a clump of values drawn together by indistinguishability, similarity, proximity or functionality (Zadeh 2006)". These characteristics, graduation and granulation, highlight the linguistic variables/functions (Zadeh 2006). Therefore, NL-computation has a pivotal role in evidence theory of modern medicine, which is basically conducted in the framework of PNL. As mentioned, routine clinical practice has been involved with *meaning-precise/value-imprecise* objects. When we consider the diabetic status of a patient with confirmed (i.e. repeated in two separated laboratory tests) fasting blood glucose (FBG) of 113 and 120 mg/dl as "pre-diabetic", we know that she/he has a biological condition which is more prone to diabetes mellitus complications than individuals with lower (<100 mg/dl) FBG; however, she/he is not diabetic in terms which intensive anti-diabetic treatments are necessary but she/he requires modifications in life-style, some medications and follow-ups. Nevertheless, we do not *precisely* know her/his FBG i.e. all clinicians know not only FBG=113 does not mean it is 113.00 but also it might be 110, 115, 117, or even 120. This reality might be better justified by considering the various uncertainties affecting medical variables (Amouzegar et al. 2014), which can be categorized into pre-analytical, analytical, and biological variation in experiments/studies (van de Ven et al. 2011). Biological variation means that the intensity of within-individual variation is proportional to the extent of measurement error, short-term biological variations (e.g. transient fluctuations, diurnal or seasonal variation), and long-life within-individual changes attributed to age, physical activity, diet, treatment, comorbidities, etc. (Amouzegar et al. 2014; van de Ven et al. 2011). These, altogether, necessitates the use of PNL in the process of evidence formation.

The concepts of precisiation and precision do have many aspects (Zadeh 2006), so that, the most important dimension of precisiation might be explained by its indexical variable ($\lambda$) whose values determine the various modalities that this process of precisiation can perform (Zadeh 2006; Zadeh 2008). $\lambda$-precisiation is mainly categorized into precision in value ($v$-precision), and precision in meaning ($m$-precision). For instance, the statement of Alex has a FBG of 116 mg/dl is $v$-imprecise and $m$-imprecise. When the repeated FBG was resulted in 113 mg/dl, he was then considered pre-diabetic. The latter is $v$-imprecise but $m$-precise. When one study says the mean ± standard deviation (SD) FBG of diabetic group was 183 ± 10, it indicates $v$-precision. However, the $v$-precision of the latter is not perfect because of inherent uncertainty of medical data, it is better to be considered as partially precise. One may suggest that fuzzification of medical variables might be a better tool for their precisiation (not only in meaning but also in value). For example, the sentence of "Alex has a FBG of 116 mg/dl" might be translated into "Alex is pre-diabetic with membership degree of 0.7". The latter approach is more practical and clinician-friendly, and the resultant is both $m$-precise and $v$-precise because not only the concept of being pre-diabetic is precise here, but also the membership degree is clear to be judged or compared with other individuals or groups. It is of particular importance to note that data compression and summarization, the major descriptive statistical tools, are forms of $v$-imprecisiation (Zadeh 2008); and are only operative in *collective* (*aggregate*) forms of data hence do not provide individualized methods that we need for clinical usage. On the contrary, using membership degrees for describing individuals enables us to employ powerful methods of inference, based on similarity measures, by comparing our patients to the studied population of evidence. For example, when one physician tries to find the best evidence for care of Alex, founds trials supporting the life style modification as a good choice for treating pre-diabetes, and finds out that Alex's *degree of similarity* with those studied pre-diabetic individuals is as high as 0.863 regarding all the inclusion/exclusion criteria and the investigated characteristics; thus the physician will be more confident to choose the approach for Alex and reassuring him.

*Fundamentum 6.* **Evidence formation optimizes the balance between homogeneity and generalizability.**

As mentioned, internal vs. external validity of study designs in the realm of EBM influences the practicality of PNL sentences in a manner which I liken it to Heisenberg's uncertainty principle in quantum mechanics (Heisenberg 1927), so the more homogeneously the study population of one study is recruited, the less precisely the findings of this study are generalizable to the routine clinical practice (fig 5). Flexibility of inclusion/exclusion criteria (low/high homogeneity of study population) and generalizability of study (low/high similarity of study population with individuals of general population) are needed to be quantified in evidence theory. Employing generalized theory of uncertainty (Zadeh 2006) and prototype resemblance theory of disease (Sadegh-Zadeh 2008) enables us to build a feasible analytical tool for this purpose, which will be introduced in results section.

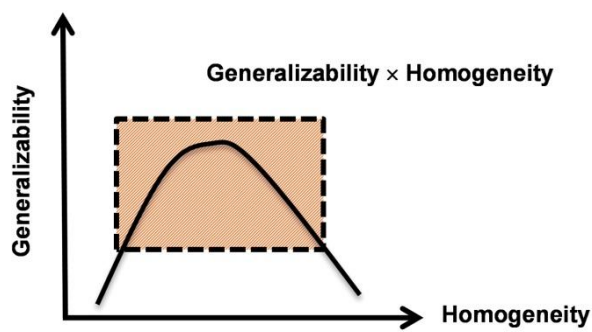

Fig 5 The relation between generalizability and homogeneity in evidence formation is displayed. An optimal balance between these two important characteristics of medical information is necessary for Precision/Personalized Evidence-Based Medicine.

*Fundamentum 7.* **Evidence formation facilitates individualized/personalized decision making.**

Despite great efforts to define and promote personalized/precision medicine (Jameson and Longo 2015), we still do not have well-defined mathematical decision making algorithms which enable us for translating the evidence of *collective* (*aggregate*) data analysis (DOE with v-imprecise statistics) to the precisely-personalized clinical decision making environment. What we need is an integrated informatics framework capable of handling individualized data in the process of evidence formation. I believe the individualized decision making is achievable when *evidence formation* is personalized itself. So it is necessary to employ identical methods of description/inference in both evidence formation and daily clinical practice environment by means of *resemblance* and *memberships* rather than naïve variability measures.

**2.2. Generalized Theory of Uncertainty (GTU)**

GTU was introduced by Zadeh LA (Zadeh 2006) based on three theses: information is a generalized constraint rather than the previous universal acceptance that information is solely statistical in nature (Zadeh 1986), everything is or is allowed to be Fuzzy (non-bivalent) (Novák et al. 2012; Zadeh 1975a; Zadeh 1975b; Zadeh 1975c; Zadeh 1975d), and information is described with natural language (Zadeh 2004). GTU was used as the pivotal method of generalized resemblance theory of evidence in this paper because it seems information of medical evidence is better handled by GTU.

In the first thesis of GTU information about X, $I(X)$ is defined as a generalized constraint on X, $GC(X)$; where X is a variable taking values in $U$. $GC(X) = X$ isr $R$, where $X$ is the constrained variable; $R$ is a constraining relation which is non-bivalent; finally r is the indexing variable that determines the modality of the constraint (i.e. its semantics) (Zadeh 2006). For example individuals in Study X (X) are pre-diabetic (r=blank, means *possibilistic*), where pre-diabetic is a non-bivalent fuzzy set (fig 6). Simply, "Alex has FBS=121 mg/dl" is a sentence with statistical nature and minimal practical importance. However, its translation as a generalized constraint, "Alex is pre-diabetic to the extent 0.63 (membership degree)" is in the natural-language usage of clinical practice.

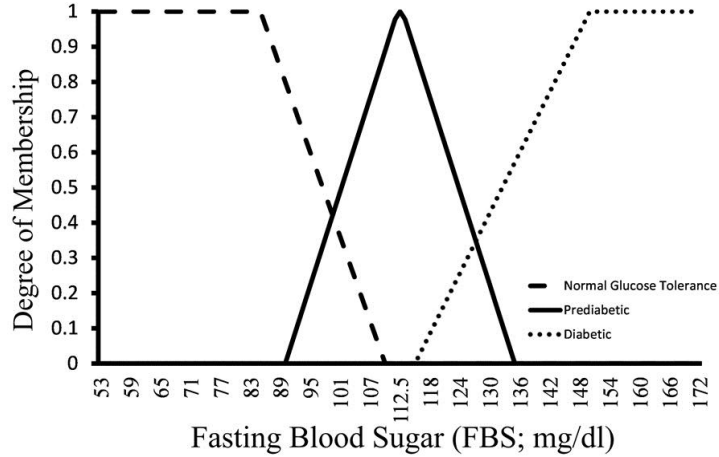

$$Pre-diabetes = \begin{cases} 0 & u < 90 \\ \dfrac{u-90}{112.5-90} & u \in [90, 112.5] \\ \dfrac{135-u}{135-112.5} & u \in [112.5, 135] \\ 0 & u > 135 \end{cases}$$

Fig 6 Membership function of pre-diabetes condition is displayed. Here, degrees of membership are based on triangular membership function

Second thesis rests upon Fuzzy sets which were introduced by Zadeh LA in the seminal paper of "Fuzzy sets" in 1965 (Zadeh 1965). Before, in 1962 he mentioned that to handle biological systems "we need a radically different kind of mathematics, the mathematics of fuzzy or cloudy quantities which are not describable in terms of probability distributions" (Wang 1997). A Fuzzy set $A$ might be expressed as $A = \{(x, \mu_A(x)) | x \in \mathbb{U}\}$ where $\mu_A(x)$ is the membership degree of element $x$ from $U$ in the set $A$.

The third thesis of GTU is PNL which is achieved by NL-computation (Zadeh 2004). First INL (Initial information set in a Natural Language) is established by a given proposition (p). Second, QNL (Query expressed in a Natural Language) is given based on a query (q). The function of GTU is to find an answer to q given p, ans(q|p), so that the deduction of ans(q|p) is involved by these modules: precisiation module, protoform module, and computation/deduction module (Zadeh 2004; Zadeh 2006). Precisiation module rests upon a mathematical operation which precisiates its operand. Here, the operand and the result of precisiation are named precisiand and precisiend, respectively (Zadeh 2004; Zadeh 2006; Zadeh 2008). The precisiation module results in a precisiand, p*. The protoform module is the mediator between the precisiation module and the computation/deduction module, which its output is a protoform of p*, abstracted as p**. The computation/deduction module is a database of *deduction rules* (Zadeh 2004; Zadeh 2006; Zadeh 2008).

### 2.3. Prototype Resemblance Theory (PRT) of Disease

PRT tries to reconstruct the concept of *disease* as a category which is not formed based on the set of prevalent features of its members (individual diseases); however it will be well-defined based on *best examples* of the category, its *prototypes*, then a human condition is considered as a disease if there is a similarity relationship (resemblance) with those prototypes (Sadegh-Zadeh 2008). Based on PRT, similarity measures are employed in the proposed theory of evidence, to enable us to modify the inclusion/exclusion criteria of medical studies and more importantly to model the process of translating evidence into clinical usage and generalizations.

## 3. Results

To reconstruct the notion of evidence formation in the realms of pEBM, I have to explain some mathematical definitions which are necessary to understand its concept. Then real examples are provided.

**Definition 1.** Conditional probability is the probability of an event A given that another event B has occurred:

$$P(A|B) = P(A \cap B)/P(B) \tag{1}$$

where $\cap$ is union operator. For example the probability of getting diabetes for pre-diabetic patients who had used metformin is obtained by (1) as:

$P(\text{diabetes}|\text{pre-diabetes} \cap \text{metformin}) =$
$= P(\text{diabetes} \cap (\text{pre-diabetes} \cap \text{metformin}))/P(\text{pre-diabetes} \cap \text{metformin})$

Simply, it indicates the proportion of pre-diabetic patients who had used metformin and developed diabetes divided by all pre-diabetic patients who underwent metformin.

**Definition 2.** Conditional possibility of $i$ elements $i=\{1,...,n\}$ in $U$ to be member of a set A given that being in another set B (Massad et al. 2003):

$$Poss(A|B) = \max_{i \in U}^{n}(min(\mu_A(i), \mu_B(i))) \tag{2}$$

For example the possibility of having diabetes ($\mu_{\text{diabetes}}(i)$) for pre-diabetic patients ($i$) who had used metformin ($\mu_{\text{pre-diabetes} \cap \text{metformin}}(i)$) is obtained by (2) as:

$$Poss(\text{diabetes}|\text{pre-diabetes} \cap \text{metformin}) == \max_{i \in U}^{n}(\min(\mu_{\text{diabetes}}(i), \mu_{\text{pre-diabetes} \cap \text{metformin}}(i)))$$

where minimum operator is applied at the patients level and the maximum operator is performed for population.

**Definition 3.** Suppose $p^*$ is a precisiand (Precisiation operator) for the membership of individuals in the subgroups of epidemiology. $P^*(D|E)$ stands for the conditional probability or conditional possibility (membership degree) of developing a given disease ($D$) if exposed to a certain risk-factor or intervention ($E$). These individuals might be *at-risk* or *doomed*. The probability/possibility of being in the second group, is obtained by $P^*(\bar{D}|E)$ conditional probability/possibility of not developing the given disease ($\bar{D}$) if exposed to the suspected risk-factor or intervention ($E$), which might be *protected* or *resistant*. The individuals of third group might be *doomed* or *protected*, in which $P^*(D|\bar{E})$ is the conditional probability/possibility of getting the disease D if not exposed to the suspected cause ($\bar{E}$). Finally, the conditional probability/possibility of individuals who might be *at-risk* or *resistant* is obtained by $P^*(\bar{D}|\bar{E})$. Various effect sizes are obtained by the operations on these probabilities/possibilities. For instance:

$$\text{Relative risk} = P^*(D|E)/P^*(\bar{D}|E) \tag{3}$$

$$\text{Odds ratio} = \left(P^*(D|E) \times P^*(\overline{D}|\overline{E})\right) / \left(P^*(\overline{D}|E) \times P^*(D|\overline{E})\right) \tag{4}$$

**Definition 4.** Prototypes or core human conditions/diseases ($P_i$) are *similar* to the extent:

$$\pi = \text{similarity}\left(\sum_{i=1}^{n} \mu_i\right) \tag{5}$$

Where $\mu_i$ is the membership degree of $P_i$ $i=\{1,…, n\}$, $\cap$ and $\cup$ are intersection and union operators, respectively.

**Definition 5.** Individuals ($H_i$) $i=\{1,…, n\}$ of a sample (study, subgroup, etc.) are *similar* to the extent:

$$\Pi_v(H_i) = \sum_{v=1}^{V} \text{similarity}\left(\sum_{i=1}^{n} \mu_v(H_i)\right) \tag{6}$$

Where $\mu_v(H_i)$ stands for membership degree of human $H_i$ for variables $v=\{1,…, V\}$.

**Definition 6.** Let $\{S_1,…, S_i\}$, be a finite set of $n≥1$ medical studies with inclusion/exclusion criteria $IEC_j$, $j=\{1,…, o\}$, which themselves are particular human conditions/diseases such as {*age ≥ 25 years, body mass index ≥24 kg/m², and a pre-diabetic FBG*}. Any member of $IEC_j$ is a prototype or core human condition/disease, which is allowed to be graduated/granulated as well as its observed data $\delta_{IEC_j}$.

**Definition 7.** Studies $S_i$ with $IEC_j$ (based on definition 6) are homogeneous for evidence formation to the extent:

$$\varepsilon_{IEC_j}(S_i) = \frac{1}{o}\sum_{j=1}^{o} \text{Similaroty}\left(\sum_{i=1}^{n} \mu_{IEC_j}(S_i)\right), \forall S_i; i = \{1,…,n\}, j = \{1,…,o\} \tag{7}$$

**Definition 8.** PNL sentence(s) obtained from evidence which is formed by studies $S_i$ with $IEC_j$ are generalizable to clinical practice for a patient/human $H$ to the extent:

$$\acute{\varepsilon} = \frac{1}{o}\sum_{j=1}^{n} \text{similarity}\left(\delta_{IEC_j}^{H}, IEC_j\right), \quad j = \{1,…,o\} \tag{8}$$

**Definition 9**. For simplicity, we assume *Dissimilarity=1-Similarity* and *non-homogeneity=1-homogeneity* (but there are monotone transformations for this purpose).

**Definition 10.** E is a precisely-personalized structure for evidence formation iff there are $I(c_i)$, $R_i$, $ES$, $\tau$, $S_j$, $\Pi_v$, $\varepsilon$, $\varepsilon'$, $p^*$, and T so that:

1. E = ⟨ $I(c_i)$, $R_i$, $ES$, $\tau$, $S_j$, $\Pi_v$, $\varepsilon$, $\varepsilon'$, $p^*$, T ⟩,

2. $I(c_i)$ is information about research question of interest $c_i$, which is a generalized constraint

3. $R_i$ is a group constraining relation which precisiates the equality/non-equality ($\lesseqgtr$) of the average effect sizes of studies ($\overline{ES_\tau}(S_j)$, where $\tau$ stands for indices of medical variables e.g. odds ratio) compared to null hypothesis about $c_i$.

4. $S_j$ are the conducted studies for $I(c_i)$, with determined effect size (ES) and internal similarity $\Pi_v$ (6).

5. $\varepsilon$ (7) is the homogeneity indicator for $S_j$

6. $\varepsilon'$ (8) indicates the generalizability of PNL sentence of evidence for a particular individual.

7. $p^*$ stands for the probabilistic/possibilistic precisiands of cardinal sources of heterogeneity

8. T represents up-to-date information of literature and timing status of formed evidence.

**Theory.** $\mathrm{E}(c_i) = I(c_i) = GC(c_i) = c_i \text{ isg } R_i \;;\; \forall i \in \mathbb{U};\; i = \{1, \dots, n\}$ (9)

$R_i \equiv \overline{ES_\tau}(S_j) \lesseqgtr \text{null } ES_\tau(c_i) \;;\; j = \{1,2,\dots,S\}, \tau = \{1,2,\dots,T\}$ (10)

$$\overline{ES_\tau}(S_j) = \frac{\sum_{j=1}^{n}\left(w_{ES_\tau(S_j)} \times ES_\tau(S_j)\right)}{\sum_{j=1}^{n} w_{ES_\tau(S_j)}}$$ (11)

The theory written here (9) sketches the concept of evidence formation in the realm of pEBM and says that evidence is formed as a generalized constraint from finite sets of studies. Where $c_i$ is a group variable of research questions of interest; $I(c_i)$ is information about $c_i$; $GC(c_i)$ is a generalized constraint on $c_i$, and $R_i$ is a group constraining relation (10) in which the effect sizes of variables $\tau$ ($ES_\tau$) and the equality/non-equality ($\lesseqgtr$) of the average effect sizes of studies ($\overline{ES_\tau}(S_j)$) (11) compared to null hypothesis of $c_i$ ($ES_\tau(c_i)$) might be of either probabilistic or possibilistic nature. Also $w_{ES_\tau(S_j)}$ is the weighted effect size of studies $S_j$. Components of precisely-personalized structure for evidence formation (definition 10) have to be collected and reported for pEBM.

Assume $c_1$ (medical research question) "Does metformin decrease the rate of conversion from pre-diabetes to diabetes ($\overline{ES_\tau}(S_j)$, in which $\tau$ is the odds ratio of diabetic event)?"

The $I(c_1)$ was collected (refer to meta-analysis (Lily and Godwin 2009)) from three RCTs (Knowler et al. 2002; Li et al. 1999; Ramachandran et al. 2006). The evidence is formed as $GC(c_1)$ which is a generalized constraint on $c_1$, indicates the group variable of results of studies examining "effect of metformin on the rate of conversion from pre-diabetes to diabetes (D=developing diabetes, M=metformin treatment)" might be explained by group constraining relations of the results of RCTs ($S_j$) in terms of:

$R_1 \equiv \overline{OR}(S_j) < 1$ (i.e. null=no effect for Metformin) (12)

$$\overline{OR}(S_j) = \frac{\sum_{j=1}^{n}\left(w_{OR(S_j)} \times OR(S_j)\right)}{\sum_{j=1}^{n} w_{w_{OR(S_j)}}} < 1$$ (13)

Here, when crisp/classical probabilistic effect sizes are considered (based on formula 4):

$$OR(S_j) = \frac{P_j(\overline{D}|\overline{M})P_j(D|M)}{P_j(\overline{D}|M)P_j(D|\overline{M})}$$ (14)

Where $P_j$ stands for conditional probabilities of studies $S_j$.

So far classical biostatistics has been employed to minimize the variation and heterogeneity of individuals/studies to provide an aggregate finding of effectiveness/non-effectiveness or association/lack-of-association regarding a medical condition/disease through hypothesis testing (12). Thus based on *inverse variance* approach of weighted effect sizes (Higgins and Thompson 2002) for crisp/classical probabilistic effect sizes:

$$w_{ES_\tau(s_j)} = \frac{1}{\text{variance}_{ES_\tau(s_j)} + \text{heterogeneity}_{ES_\tau(s_j)}} = \frac{1}{SE^2_{ES_\tau(s_j)} + v} \qquad (15)$$

Where, $SE^2_{ES_\tau(S_i)}$ stands for standard errors of effect sizes of studies, and $v$ is the correction constant for heterogeneity (Higgins and Thompson 2002) of studies:

$$v = \frac{\sum_{j=1}^{n}\left(w_{ES_\tau(s_j)} \times ES_\tau(S_j)^2\right) - \frac{\left(\sum_{j=1}^{n}\left(w_{ES_\tau(s_j)} \times ES_\tau(S_j)\right)\right)^2}{\sum_{j=1}^{n} w_{ES_\tau(s_j)}} - (S-1)}{\sum_{j=1}^{n} w_{ES_\tau(s_j)} - \frac{\left(\sum_{j=1}^{n} w_{ES_\tau(s_j)}\right)^2}{\sum_{j=1}^{n} w_{ES_\tau(s_j)}}} \qquad (16)$$

$S$ is the number of studies.

However, when possibilistic/Fuzzy effect sizes are considered (based on formula 4) for (12):

$$\text{Fuzzy } OR(S_j) = \frac{Poss_j(\overline{D}|\overline{M}) Poss_j(D|M)}{Poss_j(\overline{D}|M) Poss_j(D|\overline{M})} \qquad (17)$$

Where $Poss_j$ are conditional possibilities (Massad et al. 2003) of studies $S_j$.

Based on *inverse dissimilarities* we might provide weighted effect sizes as:

$$w_{ES_\tau(s_j)} = \frac{1}{\text{dissimilarities}_{ES_\tau(s_j)} + \text{non-homogeneities}_{ES_\tau(s_j)}} =$$
$$= \frac{1}{(1 - \Pi_{ES_\tau}(S_j)) + (1 - \varepsilon_{IEC_k}(S_j))} \qquad (18)$$

Where $\Pi_{ES_\tau(S_i)}$ and $\varepsilon_{IEC_k}(S_j)$ are obtained by (6) and (7), respectively; and based on considerations of definition 9.

**Recommendation**. It is highly desirable to adjust the conditional possibilities of fuzzy OR formula and other similar effect sizes for the internal similarities of individuals of certain subgroups of epidemiology (doomed, at-risk, resistant, and protected). For example, one may say that multiplication of $Poss_j(\overline{D}|\overline{M})$ (also the other conditional possibilities) with the internal similarity of individuals of this subgroup might be an unbiased estimation, which needs further studies.

**Example 1.** $c_1$ = {*effect of metformin on the event of conversion to diabetes in pre-diabetic population*} is analyzed based on $R_1$ using probabilistic approach. {$S_1$ (Li et al. 1999), $S_2$ (Knowler et al. 2002), $S_3$ (Ramachandran et al. 2006)} with inclusion/exclusion criteria {*age $\geq$ 25 years, body mass index $\geq$ 24 kg/m$^2$, FBG of 95 to 125 mg/dl and blood glucose of 140 to 199 mg/dl two hours after a 75-g oral glucose load, follow-up time of at least 6 months; they were excluded if they were taking medications altering glycemic status and/or having diseases known to seriously reduce life expectancy or individuals' ability to participate in the trials*} were analyzed.

NL-sentence of evidence=Metformin decreases the rate of conversion from pre-diabetes to diabetes

$$R_1 \equiv \overline{OR}(S_j) <^p 1 \text{ (null=no effect for Metformin)} \tag{12}$$

Decreased rate of conversion to diabetes (*p*<0.00001; overall odds ratio= 0.65; 95% confidence interval = 0.55-0.78)) (286 diabetes mellitus events in metformin groups vs. 392 events in placebo groups) and non-significant heterogeneity (outcome-based) of $S_i$ ($\chi^2_2$ = *2.12, p = 0.35, I$^2$ = 0.058*) (Lily and Godwin 2009) were found. However, $\delta_{IEC_j}$ were significantly (statistical/probabilistic nature) different between studies (table 1).

**Table 1.** Meta-analysis of baseline characteristics of studied populations of three randomized controlled trials evaluating the effect of metformin on glucose metabolism, insulin sensitivity and rate of conversion to diabetes in pre-diabetic people.

| Baseline Characteristics $IEC_j$ | Total weighted† averages with 95% CI | | Studies ($S_i$) | averages with 95% CI | | Difference* |
|---|---|---|---|---|---|---|
| | Placebo | Metformin | | Placebo | Metformin | |
| Age | 49.55 (48.89-50.22) | 50.14 (49.47-50.81) | Li et al 1999 | 50 (49.65-50.35) | 49 (48.56-49.44) | p<0.05 |
| | | | Knowler et al 2002 | 50.3 (49.68-50.92) | 50.9 (50.28-51.52) | |
| | | | Ramachandran et al 2006 | 45.2 (44.24-46.16) | 45.9 (44.90-46.90) | |
| Body Mass Index (BMI) | 32.90 (32.46-33.33) | 32.55 (32.11-32.99) | Li et al 1999 | 26 (25.26-26.74) | 26.4 (25.58-27.22) | p<0.05 |
| | | | Knowler et al 2002 | 34.2 (33.80-34.60) | 33.9 (33.51-34.29) | |
| | | | Ramachandran et al 2006 | 26.3 (25.68-26.92) | 25.6 (24.97-26.23) | |
| Fasting Blood Sugar (FBS) | 106.06 (105.18-106.94) | 105.49 (104.61-106.38) | Li et al 1999 | 131.51 (125.71-137.31) | 124.3 (118.77-129.83) | p<0.05 |
| | | | Knowler et al 2002 | 106.7 (106.20-107.20) | 106.5 (105.99-107.01) | |
| | | | Ramachandran et al 2006 | 99.08 (96.66-101.50) | 97.28 (94.83-99.73) | |
| 2-hours Plasma Glucose (2hPG) | 163.06 (161.79-164.33) | 163.34 (162.06-164.61) | Li et al 1999 | 162.13 (156.32-167.94) | 163.93 (158.40-169.46) | p<0.05 |
| | | | Knowler et al 2002 | 164.5 (163.48-165.52) | 165.1 (164.07-166.13) | |
| | | | Ramachandran et al 2006 | 154.93 (152.81-157.05) | 153.13 (150.99-155.27) | |
| Glycosylated hemoglobin (A1C) | 5.98 (5.94-6.02) | 5.98 (5.93-6.02) | Li et al 1999 | 7.4 (7.14-7.66) | 7.3 (7.03-7.57) | p<0.05 |
| | | | Knowler et al 2002 | 5.91 (5.88-5.94) | 5.91(5.88-5.94) | |
| | | | Ramachandran et al 2006 | 6.2 (6.12-6.28) | 6.2 (6.10-6.30) | |
| | **Between-Group difference p>0.05 | | #Global difference/heterogeneity: NA | | | |

†Total weighted crisp average $= \sum_{i=1}^{n} w_i \bar{X}_i$; where weights are $w_i \in [0,1]$ and $\sum_{i=1}^{n} w_i = 1$; here $w_1 = 0.0188$, $w_2 = 0.8356$, $w_3 = 0.1456$ based on $S_i$ inverse variances of outcomes
* non-overlapped confidence intervals reveal significant differences between studies
# Not available (NA): there are no methods for measuring global differences of variables of multiple studies because current meta-analyses measure heterogeneity based on outcome variable.

**Conclusion:** metformin treatment (850 mg twice daily, 250 mg twice or three times daily (Lily and Godwin 2009)) considerably (*35%=1-overall odds ratio*) decreases the odds of conversion to diabetes mellitus for individuals with mean (95% Confidence Intervals) of age (50.14, 49.47-50.81), BMI (32.55, 32.11-32.99), FBG (105.49, 104.61-106.38), 2hPG (163.34, 162.06-164.61), and HbA1C (5.98, 5.93-6.02).

**Example 2.** $c_1$ = {*effect of metformin on the event of conversion to diabetes in pre-diabetic population*} is analyzed based on $R_1$ using possibilistic approach. {$S_1$ (Li et al. 1999), $S_2$ (Knowler et al. 2002), $S_3$ (Ramachandran et al. 2006)} with inclusion/exclusion criteria {*Middle age adults, Overweight or Obese body mass index, Pre-diabetes Individuals, follow-up time of at least 6 months; they were excluded if they were taking medications altering glycemic status and/or having diseases known to seriously reduce life expectancy or individuals' ability to participate in the trials*}.

NL-sentence of evidence=Metformin decreases the rate of conversion from pre-diabetes to diabetes:

$$R_1 \equiv \overline{Fuzzy\ OR}(S_j) <^p 1 \text{ (i.e. null=no effect for Metformin)} \quad (19)$$

"decreased" is a group variable of information (here in $R_1$, with possibility nature aiming to maximize the homogeneity and similarity): Decreased rate of conversion to diabetes ($p<0.00001$; overall fuzzy odds ratio= 0.6; 95% confidence interval = 0.52-0.76)) and homogeneity $\varepsilon_{IEC_k}(S_j)$ of 0.518 (table 2). It should be noted here, fuzzy odds ratio (17) mentioned above was not, actually, measured due to lack of access to IPD.

**PNL sentence**: metformin treatment (Lily and Godwin 2009) (850 mg twice daily, 250 mg twice or three times daily) considerably (*40%=1-overall Fuzzy Odds Ratio=1-0.6*) decreases the odds of conversion to diabetes mellitus for pre-diabetic (membership degrees of 0.67, 0.88, 0.85 for FBG, HbA1C and 2hGP; respectively) middle age adults (membership degree=0.88) with overweight BMI (membership degree=0.80).

**Example 3.** Let us remember the example of Alex. Assume he is a middle age adult (membership degree=0.74), pre-diabetic regarding FBG (membership degree=0.94), 2hPG (membership degree=0.91), and glycosylated hemoglobin (membership degree=0.94), and also has overweight BMI (membership degree=0.86). PNL-sentence of example 2 is generalizable to Alex biomedical condition to the extent $\varepsilon'$ that might be achieved by (8) as:

$$\varepsilon' = \frac{1}{n}\sum_{j=1}^{n} similarity\left(\delta_{IEC_j}^{alex}, IEC_j\right) =$$

$$= \frac{1}{n}\sum_{j=1}^{n}\left(\frac{\min_{j=1}^{n}\left(\mu_j(alex), \overline{\mu}_j(S)\right)}{\max_{j=1}^{n}\left(\mu_j(alex), \overline{\mu}_j(S)\right)}\right) =$$

$$= \frac{1}{n}\sum_{j=1}^{n}\left(\frac{\min(0.74,0.88)}{\max(0.74,0.88)}, \frac{\min(0.94,0.67)}{\max(0.94,0.67)}, \frac{\min(0.94,0.88)}{\max(0.94,0.88)}, \frac{\min(0.91,0.85)}{\max(0.91,0.85)}, \frac{\min(0.86,0.8)}{\max(0.86,0.8)}\right)$$

$$= \frac{1}{n}\sum_{j=1}^{n}(0.84, 0.72, 0.94, 0.94, 0.94) = 0.87$$

Here, $\varepsilon'$ =0.87 indicates it is 87% *possible* for Alex that metformin therapy (850 mg twice daily, 250 mg twice or three times daily (Lily and Godwin 2009)) be able to decrease the *possibility* of conversion to diabetes to the extent 40%. Therefore, the multiplication of two possibilities (*0.87×0.4=0.348*) might enable us *to predict* the decreased *possibilistic odds* of getting diabetes for Alex. The latter index (which might be called possibility coefficient) explains the decreased *possibility* of conversion to diabetes for Alex might be 34.8% after metformin treatment.

**Table 2.** Meta-analysis of baseline characteristics of studied populations of three randomized controlled trials evaluating the effect of metformin on rate of conversion to diabetes in pre-diabetic people.

| Baseline Characteristics $IEC_j$ | Total weighted† averages of degrees of membership with 95% CI | | Studies ($S_i$) | Averages of degrees of membership with 95% CI | | Similarity* |
|---|---|---|---|---|---|---|
| | Placebo | Metformin | | Placebo | Metformin | |
| Age‡ (middle age adults) | 0.84 (0.80-0.88) | 0.88 (0.84-0.92) | Li et al 1999 | 0.87 (0.84-0.89) | 0.8 (0.77-0.83) | 0.626 |
| | | | Knowler et al 2002 | 0.89 (0.85-0.93) | 0.93 (0.89-0.97) | |
| | | | Ramachandran et al 2006 | 0.55 (0.48-0.61) | 0.59 (0.53-0.66) | |
| Body Mass Index (BMI) (Overweight§/moderately obese¶) | 0.53 (0.41-0.66) | 0.80 (0.46-0.70) | Li et al 1999 | 0.69 (0.45-0.92) | 0.57 (0.36-0.78) | 0.654 |
| | | | Knowler et al 2002 | 0.86 (0.49-0.71) | 0.51 (0.4-0.63) | |
| | | | Ramachandran et al 2006 | 0.46 (0.28-0.64) | 0.66 (0.48-0.84) | |
| Fasting Blood Sugar (FBS) Ϝ (Pre-diabetes) | 0.68 (0.65-0.72) | 0.67 (0.63-0.71) | Li et al 1999 | 0.16 (0-0.41) | 0.48 (0.23-0.72) | 0.324 |
| | | | Knowler et al 2002 | 0.74 (0.72-0.76) | 0.74 (0.71-0.76) | |
| | | | Ramachandran et al 2006 | 0.4 (0.3-0.51) | 0.32 (0.22-0.43) | |
| 2-hours Plasma Glucose (2hPG) ‖ (Pre-diabetes) | 0.86 (0.82-0.87) | 0.85 (0.83-0.88) | Li et al 1999 | 0.83 (0.7-0.96) | 0.87 (0.74-0.99) | 0.730 |
| | | | Knowler et al 2002 | 0.89 (0.86-0.9) | 0.89 (0.87-0.91) | |
| | | | Ramachandran et al 2006 | 0.67 (0.62-0.71) | 0.63 (0.58-0.67) | |
| Glycosylated hemoglobin (A1C) Y̆ (Pre-diabetes) | 0.88 (0.86-0.90) | 0.88 (0.86-0.90) | Li et al 1999 | 0.21 (0.06-0.39) | 0.27 (0.11-0.44) | 0.255 |
| | | | Knowler et al 2002 | 0.89 (0.87-0.9) | 0.88 (0.87-0.9) | |
| | | | Ramachandran et al 2006 | 0.94 (0.89-0.99) | 0.94 (0.88-0.99) | |
| | **Between Group Similarity =0.92 | | | | #Global Similarity=0.518 | |

† Total weighted fuzzy average = $\sum_{i=1}^{n} w_i \mu_i$; where weights are $w_i \in [0,1]$ and $\sum_{i=1}^{n} w_i = 1$; here $w_1 = 0.0188$, $w_2 = 0.8356$, $w_3 = 0.1456$ based on $S_i$ inverse variances of outcomes, because individual participant data were not available to achieve inverse dissimilarities of studies based on similarity of individuals of $S_i$.

* Within-variable (inclusion/exclusion criteria, IEC) similarity of studies: $\frac{1}{m}\sum_{l=1}^{m}(\cap_{i=1}^{n}\mu_{IEC} / \cup_{i=1}^{n}\mu_{IEC})$, where $n = 3$ studies ($S_i$) and $m = 2$ (placebo vs. metformin)

** Between-group (metformin vs. placebo) similarity of weighted averages of membership degrees of $IEC_j$: $\sum_{j=1}^{o} \cap_{l}^{m} \mu_{IEC_j} / \sum_{j=1}^{o} \cup_{l}^{m} \mu_{IEC_j}$ where $m = 2$ (placebo vs. metformin) and $o = 5$ criteria ($IEC_j$).

# Global between-variable similarity of studies $\varepsilon = \frac{1}{o}\sum_{j=1}^{o} \cap_{i=1}^{n} \mu_{IEC_j} / \cup_{i=1}^{n} \mu_{IEC_j}$; where $n = 3$ studies ($S_i$), and $o = 5$ criteria ($IEC_j$)

‡ Age membership function (mf) for middle age adults:
$\mu_{age}(S_i) = (\text{mean age}(S_i) - 37)/15$; where mean age $(S_i) \in [37,52]$, and $\mu_{age}(S_i) = (67 - \text{mean age}(S_i))/15$ where mean age $(S_i) \in [52,67]$

§ BMI mf for overweight: $\mu_{overweight}(S_i) = (\text{mean BMI}(S_i) - 24)/3.5$; where mean BMI $(S_i) \in [24,27.5]$, and $\mu_{overweight}(S_i) = (31 - \text{mean BMI}(S_i))/3.5$ where mean BMI $(S_i) \in [27.5,31]$

¶ BMI mf for moderately obese: $\mu_{obese}(S_i) = (\text{mean BMI}(S_i) - 29)/3.5$; where mean BMI $(S_i) \in [29,32.5]$, and $\mu_{obese}(S_i) = (36 - \text{mean BMI}(S_i))/3.5$ where mean BMI $(S_i) \in [32.5,36]$

Ϝ FBS mf for pre-diabetes: $\mu_{pre-diabetes}(S_i) = (\text{mean FBS}(S_i) - 90)/22.5$; where mean FBS $(S_i) \in [90,112.5]$, and $\mu_{pre-diabetes}(S_i) = (135 - \text{mean FBS}(S_i))/22.5$ where mean FBS $(S_i) \in [112.5,135]$

‖ 2hPG mf for pre-diabetes: $\mu_{pre-diabetes}(S_i) = (\text{mean 2hPG}(S_i) - 125)/45$; where mean 2hPG $(S_i) \in [125,170]$, and $\mu_{pre-diabetes}(S_i) = (215 - \text{mean 2hPG}(S_i))/45$ where mean 2hPG $(S_i) \in [170,215]$

Y̆ A1C mf for pre-diabetes: $\mu_{pre-diabetes}(S_i) = (\text{mean A1C}(S_i) - 4.45)/1.65$; where mean A1C $(S_i) \in [4.45,6.1]$, and $\mu_{pre-diabetes}(S_i) = (7.75 - \text{mean A1C}(S_i))/1.65$ where mean A1C $(S_i) \in [6.1,7.75]$

## 4. Discussion

It is one of the first attempts for providing a model able to explain evidence formation in PrM. Generalized Resemblance Theory of Evidence is therefore developed to provide the informatics basis for pEBM based on the previously well-developed theories of GTU (Zadeh 2006) and PRT of disease (Sadegh-Zadeh 2008). Because of substantial importance of evidence formation process and clinical decision making in PM/PrM, the logical/analytical aspects of evidence formation requires abundant attention. For summarization, table 3 is provided here to explain the characteristics of pEBM studies.

One may therefore inquire why classical probabilistic paradigm of disease-oriented EBM (*aggregate* data) is still the mainstream of medical evidence formation after decades of applying innovative soft computing methods to medicine. It is said then that, following the invention of these theories and methods, venerable efforts were done for *Fuzzification* of medical variables (Torres and Nieto 2006) and also methods of statistics (Coppi et al. 2006; Taheri 2003). This Fuzzification of methods and variables, without rigorous elucidation of the rationale behind them for all stakeholders especially for their users, clinicians, was unproductive in the realm of instrumentalist science, so added nothing more than creative applications of Fuzzy Logic in medical research history. Historically we had been told that medicine is not a field of precise mathematical equations thus medical practice was mainly based on individual experiences and judgments of experts. Nowadays, and after more than two decades of pervasive evidence-based practice clinicians talk more about update evidence of literature rather than their own expertise. Moreover, the importance of PrM is now recognized by clinicians, health systems, pharmaceutical industry, as well as by patients and policymakers (Collins and Varmus 2015; Jameson and Longo 2015); which warrants the need for those soft computing approaches.

What we will need in our discussions is the elucidation of theory (9) in terms of its capability to handle both statistical nature of medical information, and also the generalized constraint features of medical information. As mentioned, the statistical nature of medical information is mainly categorized into descriptive statistics and hypothesis testing. In both, there is an inherent way of thinking that employs a generalized way of reducing aggregate variability (inverse variance operation in weighted/pooled effect size) and heterogeneity, in order to achieve sufficient data against effect size of null hypotheses. It is clear, and of most importance to be considered, that mathematical operations of *1-heterogeneity* or *inverse variance* (or similar operations in classical methods) do not enable us concluding how much homogeneous and/or similar the investigated "studies" are; though in the best conditions we might conclude that one particular "effect size" is partially "equal" across studies in terms of raw data. As a straight development of aforesaid considerations in PM/PrM (Hamburg and Collins 2010; Jameson and Longo 2015), managing the complexities associated with the refined nosology (classification) of diseases/conditions and massive IPD of pEBM requires an informatics system capable of handling both statistical uncertainties of medical variables and its generalizability for clinical judgment. I believe direct assessment of similarity and homogeneity (comprehensive regarding all available variables of individuals, not limited to the effect size of clinical endpoint) accompanying with using membership degrees of individuals for medical conditions/diseases instead of raw data is more fruitful. This belief that we can also use crisp metric distance/similarity measures for raw data/variables of individuals/studies is logically defective because, medical conditions are Meinongian objects (Sadegh-zadeh 2000) e.g. one might have pre-diabetes and not have it at particular time, so with a FBG=120 he/she has a membership degree of 0.67 in pre-diabetes group (fig 6) and is not pre-diabetic to some extent (but not exactly as=*1-0.67*); therefore medical objects do not obey the principles of excluded middle, contradiction, and non-contradiction of classical logic (Sadegh-zadeh 2000). Thus treating medical entities and variables as Aristotelian attributes leads to erroneous conclusions that we face in clinical practice. Second, the multivariate approach for biomedical characteristics of individuals accompanying with PNL sentences of evidence, which is crucial for pEBM, is hardly achievable by those aforementioned crisp metric methods (demonstrated by low statistical power of multivariate tests with current achievable sample sizes of medical experiments). Moreover, classical regression models cannot be directly used in pEBM because the number of covariates is much larger than the number of individuals (i.e. high-dimensional characteristics e.g. genetic, biomarker, phenotypic, or psychosocial) (Ma et al. 2015).

Therefore, similarity measures (as the cornerstone of generalized resemblance theory of evidence), especially fuzzy set-theoretic measures, easily deal with multivariate comparisons (individuals with studies, individuals together, studies together, and else) that obey triangular inequality and transitivity principles (Beg and Ashraf 2009; Tversky and Gati 1982). It is clear that similarity measures which are not influenced by the number of high-dimensional characteristics or the number of individuals might be better tools for pEBM. Also PNL sentences of formed evidence by possibilistic methods are more *clinician-friendly* when natural language words e.g. middle-age, pre-diabetic, etc. are used.

Heisenberg's uncertainty principle in quantum mechanics (Heisenberg 1927) has been the other source of inspiration for me to model evidence formation of pEBM, so that its medical version might be defined as a basic limit to the precision with the certain pairs of complementary characteristics of the evidence source regarding a concept of interest in medical evidence, as *homogeneity* of study population $S_i$ and *generalizability* of evidence sentence *e*, can be maximized simultaneously. The more homogeneously the study population is recruited (without comorbidities), the less the external validity (generalizability) of the upcoming evidence will be for *real* patients, and vice versa. It is not due to our methodological inefficiencies, but an intrinsic characteristic of evidence formation of medicine. Thus, the *resemblance* part of proposed theory not only does not obligate us conducting over-homogeneous studies (high $\Pi_{ES_\tau}(S_j)$) because we have therefore take into account $\varepsilon'$ (similarity of patients with PNL sentences of studies) must be high enough to enable us to generalize PNL sentences of evidence for our *real* patients, but also encourages recruiting various individuals in terms of multiple medical characteristics if and only if these variables are well-defined.

**Table 3.** Characteristics Personalized/Precise Evidence-Based Medicine (pEBM).

| Characteristics Personalized/Precise Evidence-Based Medicine | Examples and/or explanations |
|---|---|
| A. Objectives of studies are personalized. | 1. Determining the phenotypic characteristics of pre-diabetic patients who benefit from metformin for protection against conversion to diabetes mellitus.<br>2. Determining the genes influencing the response to metformin therapy in pre-diabetic individuals.<br>etc. |
| B. The eligibility criteria of individuals are personalized. | 1. Real-world characteristics of individuals are considered e.g. to recruit patients who have comorbidities and other common characteristics with the same prevalence in general population.<br>2. Inclusion/Exclusion criteria are determined as PNL sentences e.g. middle-age overweight pre-diabetic patients, with no severe physical activity, etc. |
| C. Personalized characteristics are considered in results. | 1. Characteristics of individuals are reported as PNL sentences (e.g. patients had overweight BMI with membership degree=0.80)<br>2. Similarity of recruited individuals and their heterogeneity is determined.<br>3. Generalizability of findings to individuals is addressed.<br>4. Estimated effect sizes are adjusted for personalized heterogeneities and dissimilarities.<br>5. Differences of subgroups (at-risk, doomed, protected, and resistant) are researched.<br>6. Individual participant data (IPD) are available. |
| D. Conclusions are PNL-sentences of clinical usage | Metformin treatment (850 mg twice daily, 250 mg twice or three times daily [34]) considerably (40%) decreases the odds of conversion to diabetes mellitus for pre-diabetic (membership degrees of 0.67, 0.88, 0.85 for FBG, HbA1C and 2hGP; respectively) middle age adults (membership degree=0.88) with overweight BMI (membership degree=0.80). |
| E. Personalized characteristics are considered in medical practice as much as possible. | Alex is a middle age adult (membership degree=0.74), pre-diabetic regarding FBG (membership degree=0.94), 2hPG (membership degree=0.91), and glycosylated hemoglobin (membership degree=0.94), and also has overweight BMI (membership degree=0.86). Similarity of Alex with meta-analysis ([34]) is 87% which indicates that metformin therapy (850 mg twice daily, 250 mg twice or three times daily) is able to decrease the possibility of conversion to diabetes for Alex to the extent 34.8% (0.348=0.87×0.4(from PNL conclusion meta-analysis)). |
| F. Meta-analyses are personalized. | 1. IPD are analyzed.<br>2. Effect sizes of non-IPD studies are weighted based on their internal homogeneity and similarity of recruited individuals.<br>3. Heterogeneities of cardinal subgroups of epidemiology (at-risk, doomed, protected, and resistant) and individual similarities are adjusted in the averaged effect sizes.<br>4. Guidelines of PNL-sentences are provided for certain subgroups of studies. |
| G. Real-time clinical experiences are collected actively. | Online personalized databases are provided to collect the real-time experiences of physicians and refining PNL guidelines. |
| H. Personalized genetic inquiries are conducted. | Based on personalized information, certain subgroups (e.g. at-risk, doomed, protected, and resistant) are investigated regarding genetic characteristics influencing the outcomes. |
| I. Nosology of diseases is refined and personalized interventions are introduced. | Based on personalized data and obtained information about genetic differences of subgroups, effective precisely-personalized interventions and therapies are achievable. |
| PNL, Precisiated Natural Language | |

Time-varying nature of medical information is the other important issue that has to be addressed in evidence theory. Lots of adverse drug reactions are reported after several years of approval of therapeutic regimens. For instance, after discovering the formation of anti-drug antibodies against monoclonal antibodies (e.g. biologic medications of some rheumatologic and hematologic diseases), the previously formed evidence (in the realm of biotherapy theory) indicating advantageous therapeutic effects of such medications became questionable (Baker et al. 2010; De Groot and Scott 2007). Also, medical evidence is highly influenced by publication bias so that studies indicating a *lack of association* are not welcomed by medical journals. The misnomer, negative studies, is itself representative of this major problem. Unfortunately when the initial study discovers an important finding, the replications after that mostly try to confirm the primary result (Chan et al. 2004; Ioannidis and Trikalinos 2005; Ioannidis 2005). The selective findings reporting along with minor to major manipulations of the findings and analyses reported, are also the other misdirection of evidence formation even in clinical trials (Chan et al. 2004). Altogether, it is important for evidence theory to be dynamic over time, i.e. be able to handle the various conclusions and replication findings. Since it is very hard to model publication bias and other time-varying characteristics of medical information, it is necessary that PNL sentences of evidence to have updated supplementary explanations about the situation of medical literature, and also to provide public access to IPD. These supplements should declare the quality and quantity of performed studies about the precise question of interest, which are currently obtained through critical appraisal of the literature.

The revolution of precision/personalized medicine and its recognition by clinicians, health systems, the pharmaceutical industry, patients, and policymakers (Collins and Varmus 2015) owes to interdisciplinary efforts of theoreticians, geneticists, biologists, mathematicians, public health scholars and else. I hope my work is a genuine correction for previously ill-defined paradigms of evidence formation in the realm of pEBM and directs enough attention developing statistical methods of non-aggregate approach to get ground in the next years ahead.

## 5. Conclusion

In summary, the proposed theory of evidence formation of pEBM, generalized resemblance theory of evidence incorporates the previous well-defined soft computing methods; to provide the theoretical basis of the contemporary paradigm shift of medical practice. The theory treats medical evidence both as generalized constraints and statistical data (aggregate and non-aggregate data), which is easily translated for clinical practice environment. Also, it easily treats most of the uncertainties in medicine. The theory, here, paves the way of having precisely-personalized practice, in which both possibility measures (e.g. degrees of membership, similarity measures, etc.) and probabilistic methods (mainly hypothesis testing) need to be employed for developing future statistical/inferential methods of pEBM.

**Acknowledgements**

**Competing Interests**

I declare no conflicting interests. This study has no funding sources.